# The study of calibration for the hybrid pixel detector with single photon counting in HEPS-BPIX

*Ye Ding, Zhenjie Li, Wei Wei, Jie Zhang, Hangxu Li, Yan Zhang, Xiaolu Ji, Peng Liu, Yuanbai Chen, Kejun Zhu*

*Abstract*—The calibration process for the hybrid array pixel detector designed for High Energy Photon Source in China, we called HEPS-BPIX, is presented in this paper. Based on the threshold scanning, the relationship between energy and threshold is quantified for the threshold calibration. For the threshold trimming, the precise algorithm basing on LDAC characteristic and fast algorithm basing on LDAC scanning are proposed in this paper to study the performance of the threshold DACs which will be applied to the pixel. The threshold dispersion has been reduced from 46.28 mV without algorithm to 6.78 mV with the precise algorithm, whereas it is 7.61 mV with fast algorithm. For the temperature from 5 ℃ to 60 °C, the threshold dispersion of precise algorithm varies in the range of about 5.69 mV, whereas it is about 33.21 mV with the fast algorithm which can be re-corrected to 1.49 mV. The measurement results show that the fast algorithm could get the applicable threshold dispersion for a silicon pixel module and take a shorter time, while the precise algorithm could get better threshold dispersion, but time consuming. The temperature dependence of the silicon pixel module noise is also studied to assess the detector working status. The minimum detection energy can be reduced about 0.83 keV at a 20 °C lower temperature.

## I. INTRODUCTION

THE HEPS-BPIX, a hybrid silicon pixel array detector, has been developed for the High Energy Photon Source (HEPS) at the Institute of High Energy Physics of the Chinese Academy of Sciences (IHEP, CAS). It bases on BPIX readout chips and operates in single photon counting mode. The detection energy is designed from 8 keV to 20 keV. The total area of the prototype detector is 16 cm × 12 cm which consists of 16 modules [1].

In the single photon counting detectors, the incident X-ray photons is converted to a charge cloud by hitting the sensor. The charge cloud moves in the electric field and creates a current pulse that is proportional to the photon energy. The charge pulses are amplified and reshaped in the readout chip, and then they are digitized and counted one by one through comparing to a threshold. The effects of dark currents and readout noise can be diminished by setting a suitable threshold. The counting result is stored digitally in the pixel and can be readout quickly [2 5].

All pixels are parallel and have independent signal process cell in the readout chip. Due to the individual variation in the processes, there is a threshold dispersion which causes different response between pixels and affects the imaging results. This threshold dispersion could be minimized by trimming the threshold of the readout chips [6,7].

This paper presents the calibration process for the silicon pixel module in HEPS-BPIX detector. The relationship between energy and threshold is quantified to set the threshold. And two algorithms of threshold trimming are implemented and discussed. This paper has the following structure: section II presents the description of detector. Section 3 shows the process of threshold calibration and threshold trimming. The temperature dependencies of noise and threshold trimming are also presented in section 3. In the last, some discussions and conclusions are presented.

## II. DETECTOR DESCRIPTION

A silicon pixel module in the HEPS-BPIX consists of silicon sensor, eight readout chips and a printed circuit board (PCB). As shown in Fig. 1, the silicon sensor is connected to the 2 × 4 readout chips by bump bonding with indium, and then connected to the PCB by wire bonding [8].

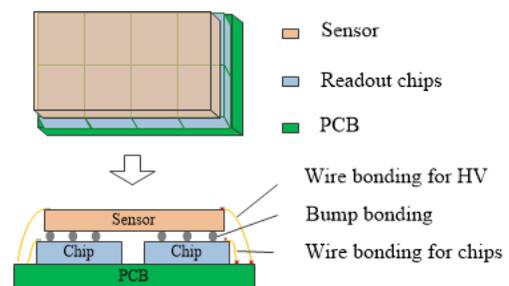

Fig. 1. The structure of a hybrid silicon pixel module in the HEPS-BPIX.

### A. The Sensor

The sensor in the silicon pixel module is fabricated on 4-inch-diameter n type silicon wafers with 300 μm thickness. A silicon pixel module has a total of 208 × 288 pixels with 150 μm × 150 μm pixel size. It consists of 8 readout chips which are designed with 104 × 72 pixels [1]. Two rows or columns of larger pixels with 150 μm × 250 μm size (3/2 of a normal pixel) span the gap of two chips. And the gaps at the vertices of four chips are covered by four large pixels with 250 μm × 250 μm size (9/4 of a normal pixel). Thus, there is no dead area for chips. For the X-ray with same energy, the count results of larger pixels are proportional to their pixel size.

The authors are with the State Key Laboratory of Particle Detection and Electronics, IHEP, CAS, UCAS, Beijing 100049, China (e-mail: dingye@ihep.ac.cn; lizj@ihep.ac.cn).

This work was supported by the Platform of Advanced Photon Source Technology R&D, PAPS.

## B. The Readout Chip

The principal properties of the readout chip in the HEPS-BPIX are listed in Table I.

TABLE I
THE PRINCIPAL PROPERTIES OF THE READOUT CHIP

| | |
|---|---|
| CMOS technology | 0.13 μm |
| Pixel format | 104 × 72 pixels |
| Pixel size | 150 μm × 150 μm |
| Dynamic range | $10^6$ (20-bit counter/pixel) |
| Clock frequency | 20MHz |
| Threshold adjustment | 5-bit DAC/pixel |

Each pixel is followed by an independent cell for processing signal. As shown in Fig. 2, a pixel cell of the readout chip is composed of an analog part and a digital part. The analog part consists of a charge sensitive amplifier (CSA), an AC coupled amplifier and a discriminator. The charges generated in the sensor are collected and pre-amplified by the charge sensitive preamplifier (CSA). The pulse signals can be injected into the CSA with a 1.6 fF capacitor to simulate the charge pulses generated in the sensor. It can be described by formula (1).

$$Q_{test} = \Delta V_{test} \times C_t \quad (1)$$

The amplitude of the pulse signals is proportional to the input energy of the sensor. Every 100mV of the pulse signals is equivalent to about 1000 e- input charge [9] . For the silicon pixel detector, each 1 keV X-rays can generate about 278 e- charges, which responds to about 30 mV of input signals. The AC coupled amplifier is used to reduce the electronic noise. Then, the pulse signal is compared to the threshold of the following discriminator. The threshold of each pixel can be set by an 8-bit global DAC (GDAC) of all pixels and its own 5-bit local DAC (LDAC). The configuration file including the parameters of the GDAC and LDAC could be configured during the test. The LDAC trimming circuit limited by the area in each pixel is non-linear and even non-monotonic. As a result of the DC differential voltage for the discriminator input, the global threshold set by GDAC should be more linear. In the digital part, the outputs of the discriminator are counted by a 20-bit counter. The configure data are latched by the latches. The 20-bit shift register is provided to configure and read data out for each pixel [1,9] .

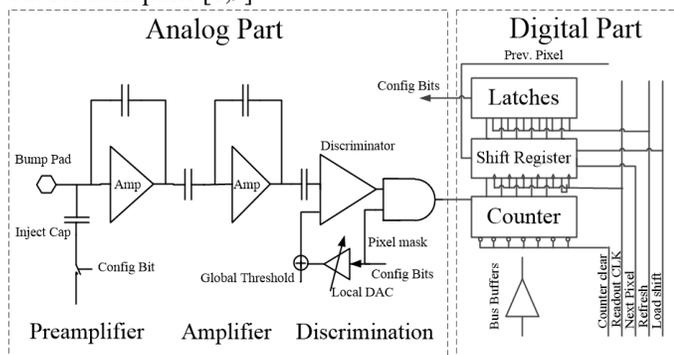

Fig. 2. The pixel signal process cell. Each pixel signal process cell of the readout chip consists of the analog part and the digital part.

The silicon pixel module is followed by a module control board and a data acquisition (DAQ) system running on the host computer to form a module detector system. The module detector system supports parallel readout of the 8 chips [10] . The larger pixel detector system can be realized by several silicon pixel modules, and an uFC control board is used to control and summarize the data of several module control boards [11] . The HEPS-BPIX 1M is composed of an array of 4 × 4 modules with one million pixels. The calibration process of the module detector system with 208 × 288 pixels is presented in this paper.

## III. CALIBRATION PROCESS

The calibration of the silicon pixel module includes the threshold calibration and the trimming for the threshold dispersion. For the threshold calibration, the quantified relationship of threshold and energy is measured to set the GDAC value responding to the input X-ray energy. For the threshold trimming, the precise algorithm basing on LDAC characteristic and the fast algorithm basing on LDAC scanning are implemented. Both of them can find the proper values of LDAC, which results in minimum threshold dispersion from pixel to pixel. In the last, the temperature dependencies of noise and threshold trimming are tested to evaluate the noise.

### A. Test Method

In the pixel signal process cell of the readout chips, we can see that the cell parameters are measured by the counting results which are related to the threshold of the discriminator. Thus, the cell parameters (i.e. offset, noise and the threshold for the input) can be tested by the threshold scanning which counts the input by changing the threshold step by step. The results of threshold scanning are shown in Fig. 3. The threshold calibration and trimming for the silicon pixel module in this paper are based on the threshold scanning of noise and input signals.

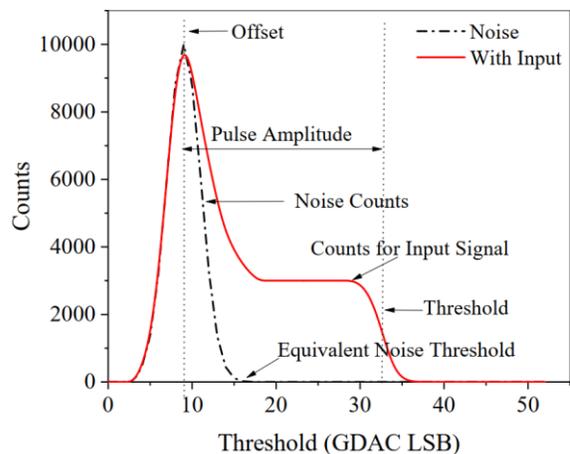

Fig. 3. Examples of threshold scanning with noise and input signals. The threshold scanning with the input signals contains the counting results for the noise as well as for the input.

*1) With noise:* the pixel noise above the threshold will be counted. In the discriminator of the pixel cell, the white noise without input should be a Gaussian distribution [12] . Take into account the offset voltage at the discriminator, the noise count rate $f_{noise}$ of a system can be described by the formula (2):

$$f_{\text{noise}}(V_{TH}) = f_0 \exp(\frac{-(V_{TH}-V_{offset})^2}{\sigma_n^2}) \quad (2)$$

where $f_0$ is the noise count rate at $V_{TH} = V_{offset}$, $V_{TH}$ is the threshold voltage of the discriminator and $\sigma_n$ is the Root Mean Square of the noise at the input of the discriminator.

A threshold is used to quantify the maximum noise of a pixel, and called the equivalent noise threshold. As shown in Fig. 3, it corresponds to the threshold of $f_{\text{noise}} = 0$. For the single photon counting detector, no noise count without input is required. Set the LDAC value above the equivalent noise threshold can reduce the noise count. Meanwhile, it determines the minimum detectable energy.

*2) With input signals:* the curve consists of two parts, one is the counting result for the noise which is similar to the curve without input, and another is the counting result for the input signal. The S-curve method is used to analyze the threshold scanning for the input signal. The pulse height from the shaper output signal is compared with the pixel threshold of the discriminator. When the threshold is lower than the height of input signals, the counting result is full count for the signals. Then the counts will decrease to zero with the threshold increasing. Due to the noise statistical fluctuation, there is a Gaussian broadened distribution when the threshold is close to the height of input signals. It can be described by the formula (3) [13]:

$$f(V_{TH}) = \frac{f_A}{2}(1-\text{erf}(\frac{V_{TH}-V_{THa}}{\sqrt{2}\sigma_n})) \quad (3)$$

where $f_A$ the is the full count rate of input signals, $V_{THa}$ is the discriminator threshold for the input signal amplitude, corresponding to the inflection point of S-curve. The standard deviation of the distribution, $\sigma_n$, is related to noise. If the input is X ray, the $f_A$ is influenced by the flux of the source and the $\sigma_n$ is also related to the charge sharing and energy spectrum of the incident X-rays. The threshold responding to 50% of the pulse signal or energy is regarded as the most efficient threshold which is effective to reduce the effect of noise as well as charge sharing [14].

### B. Threshold calibration

The threshold voltage of the discriminator of each pixel can be set by 8-bit global DAC (GDAC) shared by the same readout chip and its own 5-bit local DAC (LDAC). Usually, the GDAC is set to respond to the energy for all pixels in a readout chip, and the LDAC is used to minimize the threshold dispersion [13-16]. Therefore, it is necessary to quantify the relationship of the GDAC values and the energies.

The quantified relationship of the GDAC and energy can be acquired by scanning the GDAC values with different energies at same LDAC value. This experiment is done at the 1W2B beam station of Beijing Synchrotron Radiation Facility (BSRF) with an energy range from 6 keV to 20 keV. It provides the monochromatic and uniform X-ray by scattering. But the flux of the X-ray source varies with energies. The curves are obtained by setting the local threshold of all pixels to the middle LDAC value 15 and scanning the GDAC values repeatedly with the energies from 6 keV to 14 keV. Fig. 4 shows the threshold scans of a pixel for the noise and different X-ray energies.

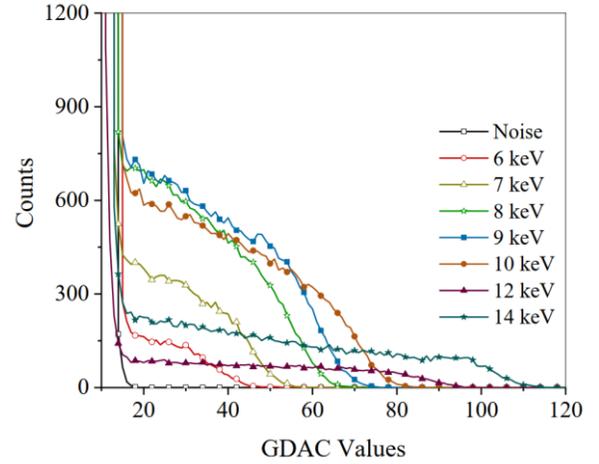

Fig. 4. The threshold scans of a pixel for different X-ray energies. The full counts influenced by the fluctuation of X ray flux vary with different energies.

The equivalent noise threshold acquired by scanning different GDAC values with noise can be used to separate the part of noise counts and the part of counts for input energy. A 7th polynomial is used to fit the S-curve for the input counts part. The $V_{THa}$ corresponding to the inflection point of S-curve can be calculated by derivation. It is the GDAC responds to the input energy.

The average GDAC of all pixels with same energy is the GDAC of chips responding to this energy. The quantified relationship of the GDAC and energy is shown in the Fig. 5, and it can be described by the formula (4).

$$\text{GDAC} = 8.43\ E - 26.54 \quad (4)$$

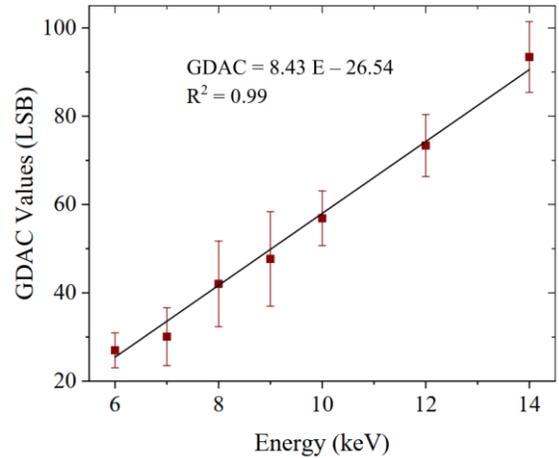

Fig. 5. The relationship of the energy and threshold. It is linear and about 8.43 GDAC LSB responds to 1keV.

It is linear and about 8.43 GDAC LSB responds to 1keV. The average of the equivalent noise threshold is 17.19 GDAC LSB. Therefore, the minimum detection energy is about 5.19 keV. For the detectable energy, the GDAC can be set to the corresponding energy.

### C. Threshold trimming

The threshold voltage adjusted by GDAC values is global for a whole chip. There is a threshold dispersion from pixel to pixel with a same GDAC value for all pixels. The purpose of threshold trimming is to find the proper LDAC values to minimize the threshold dispersion. In the readout chip, there is a 5-bit LDAC for each pixel, which means only 32 values can be set. According to the σ at the middle LDAC of value 15 for each energy, 3σ is about 60 GDAC LSB. Setting the LSB value of LDAC as twice as GDAC LSB can cover the dispersion. Theoretically, the threshold dispersion can be reduced by transferring the GDAC to the LDAC for each pixel without additional tests. However, limited by the pixel area of the readout chip, the LDAC values are not linear and even not monotonic. This algorithm is not adapted to the situation with the non-linear of LDAC which can cause another threshold dispersion. Therefore, two algorithms with necessary tests are implemented and discussed to find proper LDAC values. All the experiments of threshold trimming are tested with electronic input signals.

*1) Precise algorithm:* the LDAC characterizations of each pixel, which can describe the behavior characters of the trimming circuit, are used to find proper LDAC values to minimize the threshold dispersion.

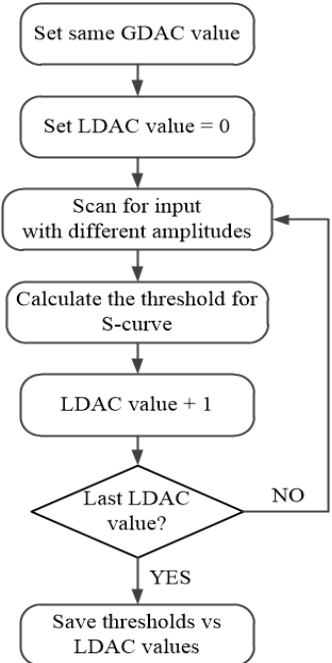

Fig. 6. The procedures of the precise algorithm for LDAC characteristic measurement.

As shown in Fig. 6, setting a same GDAC for all pixels, different amplitudes of input signal are scanned repeatedly with different LDAC values. The completed S-curve for each LDAC are obtained by repeat scanning for the input signal in a certain range from 60 mV to 600 mV. The 32 S-curves of a pixel are shown in Fig. 7 (a). Based on the S-curve method, the characteristic curves are acquired. As shown in Fig. 7 (b), there are 32 $V_{THa}$ responding to 32 LDAC values for each pixel. Averaging thresholds for all pixels, setting the LDAC value at which the threshold is closest to the average threshold for the pixel can minimize the threshold dispersion of pixel to pixel. The result is shown in Fig. 8. The threshold dispersion has been reduced from 46.28 mV without algorithm to 6.78 mV with the precise algorithm. Scanning different amplitudes with the GDAC and the trimmed LDAC values, the Equivalent Noise Charge (ENC) and threshold are shown in Table II.

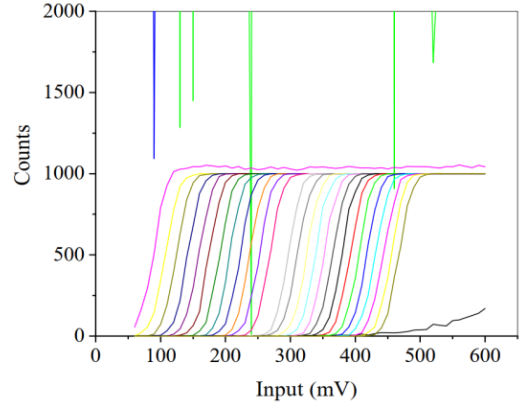

(a)

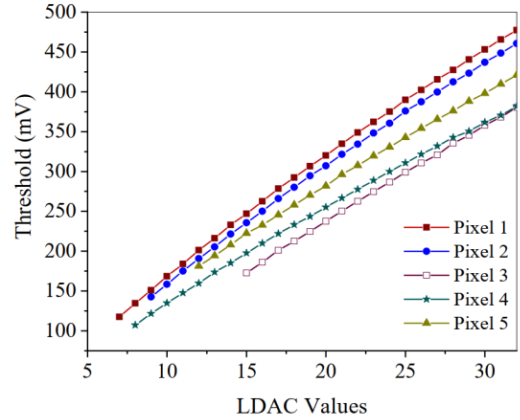

(b)

Fig. 7. The precise algorithm for LDAC characteristic measurement: (a) The 32 S-curves of a pixel. (b) The threshold-LDAC characterization curves of five pixels.

TABLE II
THE ENC AND THRESHOLD WITH PRECISE ALGORITHM.

|  | ENC | | Threshold | |
| --- | --- | --- | --- | --- |
|  | Mean/e- | δ/e- | Mean/mV | δ/mV |
| Untrimmed | 171.34 | 22.348 | 255.47 | 46.28 |
| Precise algorithm | 162.25 | 18.030 | 263.56 | 6.78 |

This threshold trimming is a time-consuming process. For each LDAC value, scanning the 55 amplitudes of input signals needs 1760 steps. The test time is about 147 minutes with 5s per step. Added the calculating time which varies due to the algorithm processor speed, the total time for LDAC characteristics tests of a silicon pixel module is about 3 hours which would consume more time with a smaller step.

IEEE TRANSACTIONS ON NUCLEAR SCIENCE

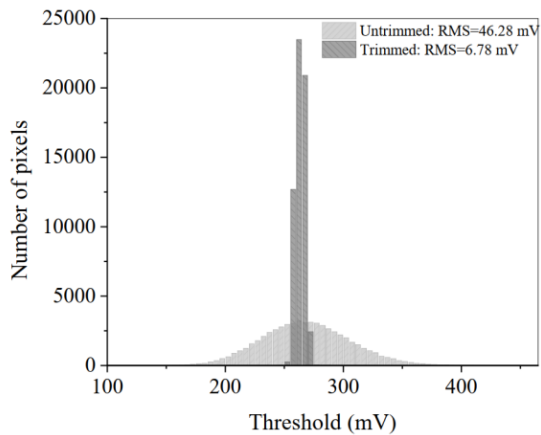

Fig. 8. The threshold trimming result with the precise algorithm. For the untrimmed, the LDAC values of all pixels are 15.

*2) Fast algorithm:* for finding proper LDAC values, the time-consuming LDAC characteristics tests are possible to be omitted by the just LDAC scanning with a fixed input signal.

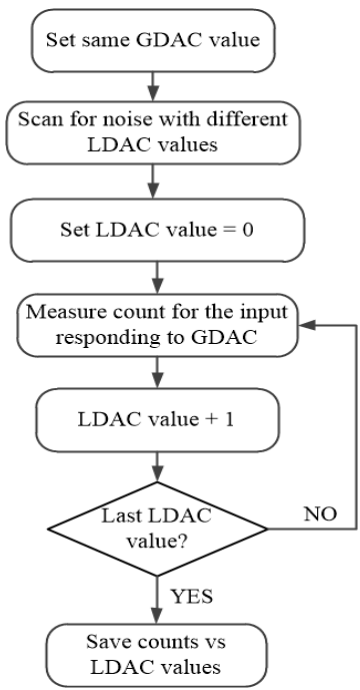

Fig. 9. The procedures of the fast algorithm.

As shown in Fig. 9, setting a same GDAC for all pixels, the equivalent noise threshold can be obtained by scanning for noise with different LDAC values. Then, fixed the input amplitude corresponding to an energy, scanning the LDAC from 1 to 32, each pixel can get a S-curve. The scanning curves of 59904 pixels for the noise and a fixed input are shown in Fig. 10. Removing the counts for noise and fitting the counts for input to the S-curve, the LDAC corresponding to the inflection point of each S-curve is the LDAC threshold of each pixel.

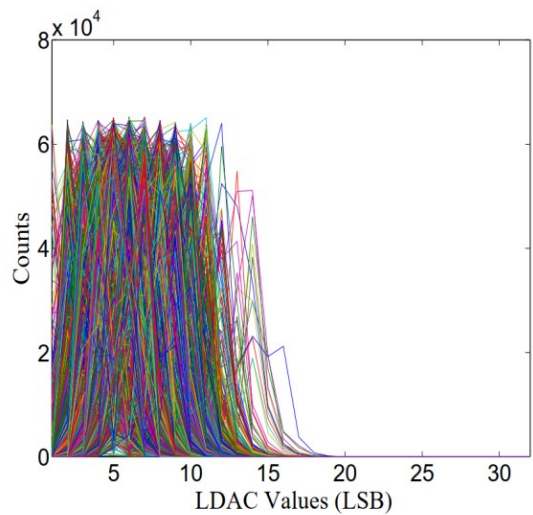

(a)

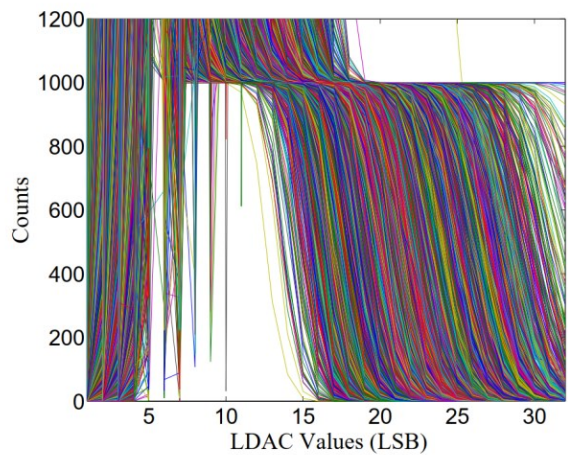

(b)

Fig. 10. The fast algorithm: (a) The scanning curves of 59904 pixels with noise. (b) The scanning curves of 59904 pixels with a fixed input.

Scanning different amplitudes with the GDAC values and the trimmed LDAC, the result of threshold trimming is shown in Table III and Fig. 11. The threshold dispersion has been reduced from 46.28 mV without algorithm to 7.61 mV with the fast algorithm.

TABLE III
THE ENC AND THRESHOLD AFTER THRESHOLD TRIMMING.

|  | ENC |  | Threshold |  |
|---|---|---|---|---|
|  | Mean/e- | δ/e- | Mean/mV | δ/mV |
| Untrimmed | 171.34 | 22.348 | 255.47 | 46.28 |
| Fast algorithm | 162.21 | 49.170 | 266.12 | 7.605 |

This algorithm is similar to do a flat field for all pixels, and the noise can be reduced by setting the LDAC above the equivalent noise threshold. It can short the test time by just scanning 32 LDAC values for noise and an input pulse signal. The test time is about 6 minutes with 64 steps and 5 seconds per step. Added the time of simple calculation, the total time of a silicon pixel module is about 8 minutes.

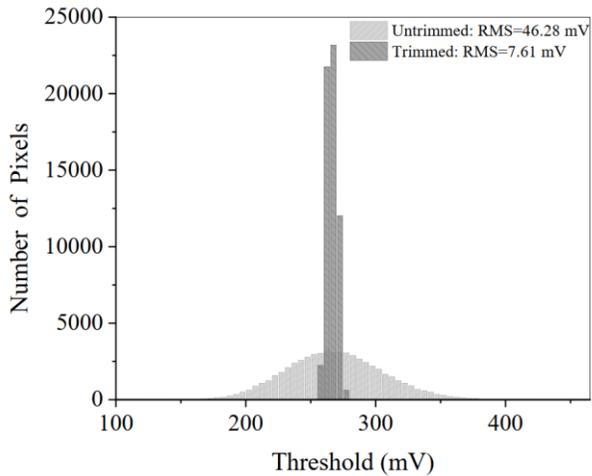

Fig. 11. The threshold trimming result with the fast algorithm. For the untrimmed, the LDAC values of all pixels are 15.

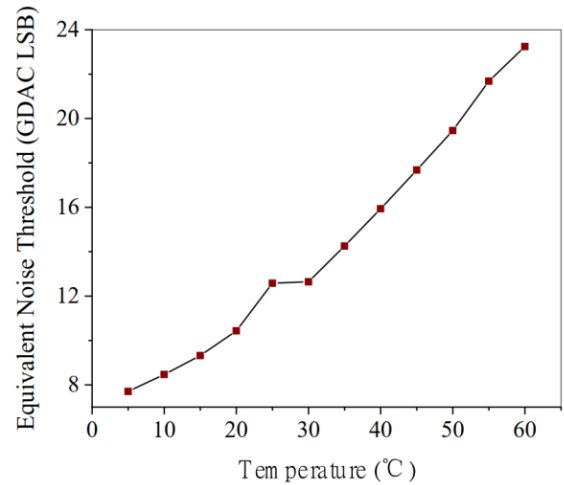

Fig. 12. Temperature dependence of the equivalent noise threshold.

### D. Temperature dependence

All the tests of threshold calibration and threshold trimming were operated at room temperature above. The pixel process cell in the readout chips consists of many transistors whose many parameters such as threshold voltage, mobility, the intrinsic carrier concentration of silicon, etc, vary with temperature [17, 18]. Therefore, we tested the stability of the equivalent noise threshold and threshold trimming vs temperature changing from 5 °C to 60 °C with a step of 5 °C.

As shown in Fig.12, the equivalent noise thresholds at each temperature are obtained by setting the middle LDAC value 15 and scanning different GDAC values for noise. The equivalent noise threshold rises with the increasing temperature. Compared with the room temperature 25 °C, the equivalent noise thresholds at 5 °C has decreased about 7 GDAC LSB, which means that the minimum detection energy can be reduced about 0.83 keV. As shown in Fig. 13, the ENC and the threshold dispersion rise with the increasing temperature. With the trimmed LDAC values at room temperature 25 °C, the change in the threshold dispersion of precise algorithm is in the range of about 5.69 mV, while it is about 33.21 mV for the fast algorithm. After the re-correction with the fast algorithm at each temperature, the change in threshold dispersion of fast algorithm is about 1.49 mV.

### E. X-ray imaging

The silicon pixel modules after the calibration imaged at the 1W2B beamline of BSRF. The imaging results for the powder diffraction ring of LaB6 sample at 12 keV X-ray are shown in Fig. 14. Analyzing the imaging data of the silicon pixel module, Fig. 15 shows the diffraction angles and the relative intensity of each diffraction ring. The four Bragg diffraction angles of the LaB6 sample are 0.26 rad, 0.36 rad, 0.44 rad, and 0.51 rad.

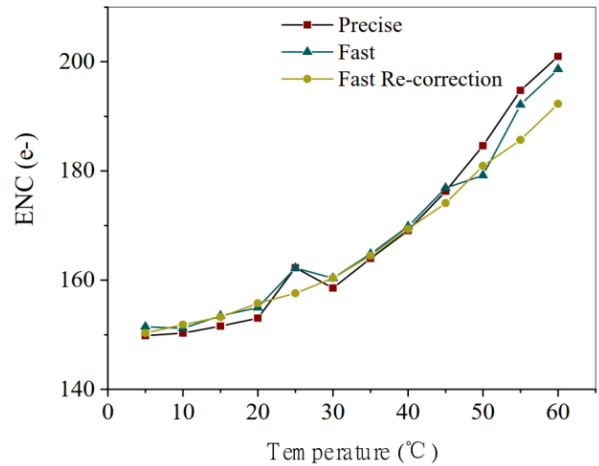

(a)

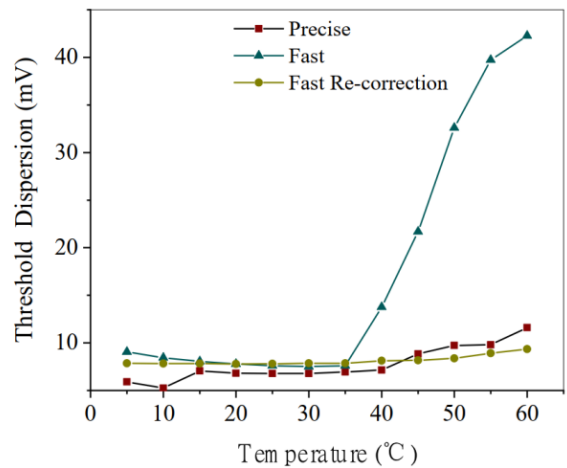

(b)

Fig. 13. Temperature dependence of threshold trimming: (a) ENC. (b) Threshold dispersion.

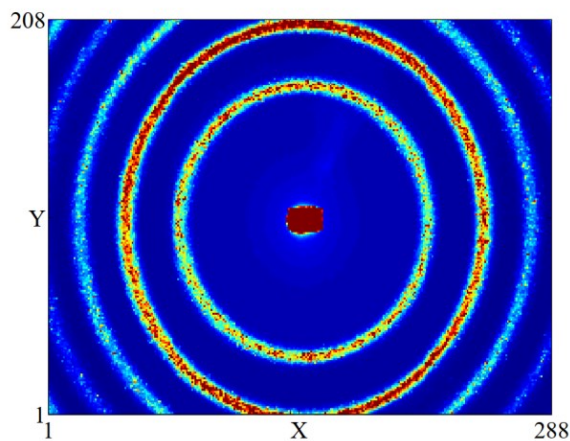

Fig. 14. The imaging results of the LaB6 sample's powder diffraction.

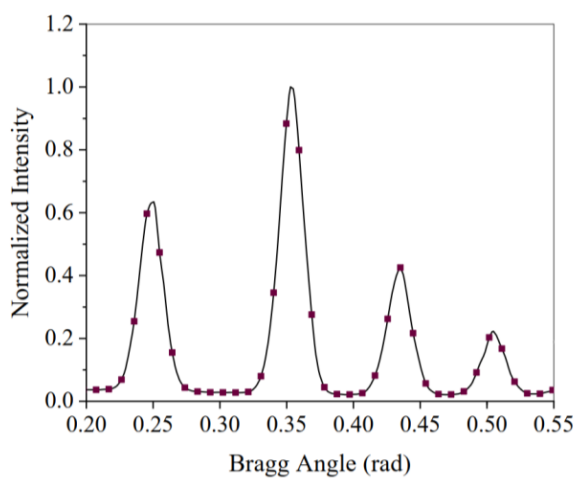

Fig. 15. The diffraction angles of the diffraction ring. Each diffraction angle responds to a diffraction ring in Fig. 14.

## IV. DISCUSSION AND CONCLUSION

This paper presents the calibration process for the silicon pixel module in the HEPS-BPIX detector system, which includes threshold calibration and threshold trimming for the threshold dispersion. For the threshold calibration, the relationship of the energy and threshold is quantified, which is about 8.43 GDAC LSB for a 1 keV energy. The minimum detection energy is about 5.19 keV with an equivalent noise threshold of 17.19 GDAC LSB. In the range of detectable energy, the GDAC can be set to the corresponding energy. For the threshold trimming, the threshold dispersion has been reduced from 46.28 mV without algorithm to 6.78 mV with the precise algorithm, whereas it is 7.61 mV with fast algorithm. With the trimmed LDAC values at room temperature 25°C, the threshold dispersion of precise algorithm the changes in the range of about 5.69 mV, while it is about 33.21 mV for the fast algorithm. After the re-correction with the fast algorithm at each temperature, the change in threshold dispersion of fast algorithm is about 1.49 mV. The measurement results show that the precise algorithm can get better threshold dispersion with the LDAC characterizations for a silicon pixel module, but it is time-consuming, whereas the fast algorithm can short time by LDAC scanning and get an acceptable result. The fast algorithm is effective and well adapted to re-correction of temperature and even the calibration with the X-ray.

The temperature dependence of the silicon pixel module noise is also studied. The equivalent noise threshold rises with the increasing temperature, the minimum detection energy can be reduced about 0.83 keV at a 20 °C lower temperature. Working in a low temperature is important for a good detector status. In addition, the silicon pixel module after the calibration has been applied to the measurement of X-ray diffraction ring. In the future, more studies about the calibration of threshold trimming with X-ray will be performed and compared with electronic input signals.